# Why dream? A Conjecture on Dreaming

Marshall Stoneham
London Centre for Nanotechnology
University College London
Gower Street, London WC1E 6BT, UK

**Abstract** I propose that the need for sleep and the occurrence of dreams are intimately linked to the physical processes underlying the continuing replacement of cells and renewal of biomolecules during the lives of higher organisms. Since one major function of our brains is to enable us to react to attack, there must be truly compelling reasons for us to take them off-line for extended periods, as during sleep. I suggest that most replacements occur during sleep. Dreams are suggested as part of the means by which the new neural circuitry (I use the word in an informal sense) is checked.

## 1. Introduction

Sleep is a remarkably universal phenomenon among higher organisms. Indeed, many of them would die without sleep. Whereas hibernation, a distinct phenomenon, is readily understood in terms of energy balance, the reasons for sleep and for associated dreaming are by no means fully established. Indeed, some of the standard ideas are arguably implausible [1]. This present note describes a conjecture as to the role of dreaming. The ideas, in turn, point to a further conjecture about the need for sleep.

I have tried to avoid any controversial assumptions. All I assume about the brain is that it comprises a complex and (at least partly) integrated network of components, and has several functions, including the obvious ones of control of motion, analysis of signals from sensors, computing (in the broadest sense), memory, and possibly other functions. Some brain components may provide a degree of redundancy; others may have more subtle functions not yet defined. I do not try to identify detailed physical mechanisms. In particular, my conjectures do not need an understanding of the means by which information from DNA or elsewhere mediates or controls brain structures or their replication.

## 2. Background .

Most cells in the body are replaced every few months. This seems the case for stem cells [2] and for fat cells [3]. Neurogenesis, the creation of new neurons, is established for neurons that migrate to the olfactory bulb and also for those in the dentate gyrus of the hippocampus associated with spatial memory [4]. There are exceptions, e.g., some eye structures and some iron-containing molecules. Proteins within neurons are replaced with half-lives from hours to weeks, but (private communication, Luca Turin) many, perhaps most, neurons are never replaced throughout life. Likewise, most biomolecules are replaced on similar time scales. Thus basic molecular constituents of some components of the brain - memory and networks of whatever nature - are largely replaced every few months. As regards memory, irrespective of whether it is maintained physically in molecular or cell-level structures, there has to be regular, perhaps continuous, refreshment of some of the relevant molecular hardware.

Related to these ideas of physical replacement, I note two important observations. First, there is evidence that those nerve connections that have built up during the working day are cut back to some degree during sleep, partly to edit information and partly to reduce energy use [5].  Secondly, during sleep, certain categories of genes are enhanced. These prove to be those that are involved in rebuilding the molecules that are essential for cell function. Thus the enhancement affects genes for all stages of cholesterol synthesis. Cholesterol has a role in signalling, since it is a major component of the lipid rafts in cellular membranes that bring together signalling molecules such as neurotransmitters and receptors, and so affect the efficiency of signalling [6]. This is consistent with conjecture A below that much of the  regular rebuilding of the neural system occurs primarily during sleep.

The renewing brain surely makes use of any spare capacity and the redundancies. Spare capacity seems essential to the human brain. Not surprisingly, the brain has evolved to have robust neural circuitry, rather than maximally efficient circuitry: lean management (that fashion of managers in industry) is not an option. The relevant parts of the brain for present purposes are presumably those showing neural correlates of consciousness, of which there are quite a few [7]. Lesion studies suggest there is not much redundancy evident in the hindbrain or midbrain, though it might be there at smaller length scales that are hard to identify. But there does seem that quite a lot of the cortex can be destroyed without loss of function, and thus offering some redundancy. It is the cortex where most of the brain activities occur concerning seeing, hearing, language and movement. Adjacent to it is the claustrum which, with its neural connections to most parts of the cortex, has been suggested by Crick and Koch as having a central role for consciousness [8]. Recognition of the continuing renewal of brain components (including structures in the claustrum and the cortex, and presumably also the "memory biochemicals" [9] has several consequences.

3.     Consequences of Replacement and Renewal

**3.1 Scheduling and Validating changes in neural circuitry**     The replacements of active components of the brain, whether cells or molecules, has to be done in a way that does not interfere with normal functions in an unacceptable manner. After all, one important brain function is being able to move away from dangers. It would not be good to run out of brain function when pursued by a tiger. However, we do sleep, and we do so at times and places over which we (usually) have some control, such that we are not exposed to these dangers. So I suggest Conjecture A:

**Conjecture A**     The regular replacement of basic brain components takes place primarily when we sleep, and indeed sleep is necessary to allow these physical replacements to be achieved with minimum danger from external events.

During deep non-REM sleep, neuronal firing cannot spread across the cortex [10], as one would expect if maintenance activities were in progress, rather like rail network repair, when engineering work is done at night. The cortex has reserve circuitry such that most, if not all, faults can be repaired "off line" during sleep [1 p 61]. Sweating occurs more during sleep [1 p 34], possibly consistent with molecular replacement activity. Indeed [1 p 54] cell division typically has a peak around 2 am. However [1 p 53], it seems mainly a misconception that sleep aids general body recovery: sleep is essential for the cortex, but does not seem to have much to do for the rest of the body, though complete sleep deprivation leads rapidly to death in experimental animals [1, 9]. Hibernation is not equivalent to sleep [1 p 9]: hibernating animals need to warm up before they can sleep, so perhaps molecular and cell replacements do not occur significantly during hibernation.

Any replacements of neural circuitry has to be checked: does it function correctly? I propose:

**Conjecture B**, that dreams are part of the brain's way of checking that the neural circuits and memory stores are working correctly.

The natural time to make such checks is when the period of physical replacement is essentially complete, i.e., at around the time of awakening, which is indeed when many dreams occur. I do not rule out the possibility that sleep and dreaming may permit tidying up the information in the brain, but suggest tidying is not a main objective.

There might be a biochemical indication that triggers dreams and awakening when most of the relevant physical changes are over. Hobson and McCarley [12] propose an Activation-Synthesis hypothesis, postulating a "non-specific state generator" that leads to activation (see especially their figure 10). No origin for the non-specific state generator seems to be suggested, and indeed it could be purely stochastic in their model, though they do make the point (item 1 on page 1346) that the primary motivating force for dreaming is physiological, not psychological. My conjecture B suggests one *specific* physiological reason and links it to a specific physiological reason for sleep, namely my conjecture A. There may be further reasons. Certainly, there are links to other physical phenomena. During a dream, actions associated with what one dreams may be inhibited. During REM (rapid eye movement) sleep, a whole group of molecules involved in nerve transmission is shut down [13] "which is why when you dream of running you don't kick your partner to death"). If one wants a computer analogy, one cannot use a program like WORD while booting up a computer. There may well be ways that molecules like tyramine influence checks of the memory/brain system, as well as inhibiting actions. However, there seem strong arguments that dreaming and REM sleep are controlled by different brain mechanisms [16]. Hobson [12, 15] emphasises that dreams can indeed occur during non-REM sleep, but that they are far more likely to occur during REM sleep.

4.     **Dreams and their content**

Checks of neural function will have some bias towards things that are genuinely important, and probably to events that are recent. Regarding the bias to the *important,* if (see below) the *main* use for a brain, at least for simpler animals, is to enable evasive action when threatened, then perhaps it is not surprising [17] that 32% of nightmares concern running away, nearly 5 times more than the next most quoted topic.

Checking is not the same as using, so it is no surprise that dreams link items apparently irrationally. The important and the irrational may get mixed up. Thus it has been said - I think by Fritz Spiegel (in an unverified letter to *The Times*) - that if, when running away in a dream, one gets stuck in mud, one should unscrew one's legs at the knees, remembering that the left leg has a left-hand thread.

Brain functions include links to at least some of the sense organs. Perhaps the rapid eye movements (REM) are (in part) associated with checks of links to the visual system. Whilst most dreams seem visual, dreams can involve any senses. Discussions with friends indicate that most, like me, can remember several dreams that have sound as well as colour. Very few seem to dream odours. One exception is scent scientist Luca Turin (private communication 2007) who reported that he dreamt of an actual fragrance that he was commenting on in his dream. Other exceptions are war veterans who [18] have exceptionally bad and realistic dreams, in which they smell "blood and barbequed flesh."

Both during and after sleep deprivation [1 p 37] daydreams are more prevalent; the Table of [1 p 45] of experiences include "misperceptions" rather than "hallucinations." This fits with Conjecture B, in that misperceptions would correspond to neural circuit testing mixing the realities of the senses with past experiences.

Freud felt that dreams carried deep and significant messages about his patients, and that these messages could be brought from hiding by methods such as free association. Others doubt that dreams carry emotional messages at all [16]. Still others have ascribed dreams to brain chemistry, though not always defining the nature or value of that chemistry. My conjecture implies that there may be emotional messages, but these could be irrationally mixed with other signals, and so full of potential ambiguity. The conjecture would certainly suggest that dreams could be associated with parts of the memory (and of the brain system as a whole) that, when awake, the patient might not be aware of, or might wish not to access, or would feel inhibited from accessing. But it would be wrong to take the messages as either significant or systematic. If you were diverted off a motorway, you will explore country and see places you did not anticipate, but what you do see will not define your inclinations or your experience, and you will leave lots of countryside unexplored.

## 5    Dreams and physical damage

One might expect unusual dreams for people whose brains have suffered major physical damage. So could there be a pathology of dreams? When particular parts of the brain are disabled, should there be particular dream patterns (rates, intensity, subject matter) as the brain recovers to a state that can be usefully checked? A recent report of a brain-damaged woman who heard voices, these all having the same speech impediment that she had developed [19].

Conjecture B suggests that there would be more dream events the more crucial the role of new (replaced) neurons. Imayashi et al [4] investigated two cases of neurogenesis. In the first, they found neurons in the olfactory bulb died regularly, and were replaced, but - even when replacement was stopped - there were no deficits in smell memory. This suggests little need to check the olfactory neural circuitry since it is robust under neuron death and genesis, and indeed my informal enquiries suggest that very few people indeed have dreams involving scents or smells. Imayashi et al also found neurons in the dentate gyrus of the hippocampus do not seem to die regularly, yet stopping neurogenesis for these neurons did interfere with spatial memory. Again, from my informal enquiries, it seems to be the case that people do have dreams that involve irrational links between places and settings, i.e., where spatial memory might be involved. For instance, I myself have had a number of dreams that mix familiar and wholly imaginary landscapes, as could happen in the checking of the related neural circuitry. The observations of [4] again are consistent with the realities of the senses getting mixed by neural circuit testing to past experiences.

## 6    Further conjectures

Conjectures A and B can be supported by evidence that, whilst not compelling, does make contact with previous work and previous experiments. I now add two conjectures that are much more speculative.

**Conjecture C:** Consciousness includes the mechanism that checks and validates the interconnections within the brain.

The first reason for this suggestion is that, as information processing becomes more complex, there will come a time at which there is a strong evolutionary advantage to self-check. Daniel Wolpert [20] observed that the brain is needed to guide evasive movements, and noted that the sea anenome [I think that was his example] has a central nervous system for the stage of its life when it is mobile, but loses it once fixed to a rock and immobile. Wolpert argued that the brain uses Bayesian methods to control motion when there is inexactitude in the system: there is noise in the input signals, and anyway we cannot move ourselves with the ideal levels of precision that we might like. The simplest evolving brains (if I have rightly understood Wolpert's ideas) would be capable of Bayesian analysis, with a continual update of how to respond to changing events. Those forms of life with spare capacity would have the opportunity to gain further evolutionary advantage through gradual upgrades of this Bayesian system. At some critical level, there becomes a need to check for malfunctions and find ways round them. Even the simplest PC needs rebooting regularly, whereas you don't need anything special when you just count using your fingers. Such malfunction checks would include consistency checks for the different senses. I am indebted to Dr Luca Turin who, on reading an early draft, gave a nice analogy of consciousness as "among other things, a sort of instrument panel that tells us whether there are discrepant readings indicating brain malfunction…for example the dreadful illness and nausea that arise when balance organ input differs from visual input, as in seasickness."

Secondly, consciousness is associated with certain forms of creativity. For example, (as Chomsky noted [21]) we routinely produce sentences that we have never said before. Indeed, we may alter a sentence mid-way because of the response we are getting. Dreams and unconscious thought (in the informal, lay sense, e.g., thoughts about topic X when our concentration is on topic Y) can sometimes be creative. Koestler [22], in a book whose main thesis was thoroughly mauled by Medawar [23], made the hypothesis (vastly simplified) that a previously unexpected juxtaposition of ideas from significantly different backgrounds, gave rise to laughter, or discovery and synthesis. This led Koestler and Medawar to various arguments about whether dreams were deep private messages or whether they actually had no meaning at all. I will not enter into the Koestler-Medawar argument (Medawar's objections to Koestler are surely right but do not seem to apply to what I suggest), but merely note that my Conjecture B would naturally lead to unexpected juxtapositions of ideas or images when dreaming. In other words, during renewal and checking of brain components, some of the previous connections will not be available, and new ones will be substituted, at least temporarily. Most links will be dross. A few of the new links might be recognised somehow as "useful" and either creative or funny (Kekulé was lucky). If the brain has spare system capacity, these novel juxtapositions might simply be stored for some time. Some will just disappear as brain components get replaced. Others, perhaps, may survive long enough for the useful ones to give evolutionary advantage. These other juxtapositions might be complex, e.g., enabling non-algorithmic mathematics. Just how the brain recognises the useful new linkages as valuable I do not know, though perhaps there are useful clues in the paper by Snider [24].

**Conjecture D:** Qualia correspond to the colour-coding of hard wiring in a computer, i.e., the qualia are not necessary (people can live happily with colour blindness), but the qualia make things easier, including in checking the brain connectivities and the basic functioning of the memory. The qualia thus give a simple aid in quickly scanning some part of brain operation. Again, I am indebted to Dr Luca Turin for the observation that "The fact that hallucinogens *increase* our perceptions is a sure sign that under normal circumstances they are kept to a minimum."

## 7     Conclusions

The understanding of dreaming does not lack its frustrations. The medieval Latin lyric *Dreams, dreams that mock us with their flitting shadows* [25] has its modern parallels. Yet sleep and dreaming are so common that they surely have an identifiable physiological purposes. Moreover, these must be both basic and fill a need at all stages of the lives of higher animals. My two conjectures, right or wrong, satisfy those minimal conditions: sleep offers a period for physical renewal through routine replacements of cells and biomolecules at times that minimise vulnerability, and dreams then offer a means by which new neural circuitry be checked. I would not wish to underplay other hypotheses, notably the traditional view that memories are replayed and reinforced during sleep [26, 27], but such ideas do not seem sufficient to explain either why the brain needs many of its functions to be switched off for so long each day, nor many features of dreams, such as their irrational mixtures of memory fragments.

**Acknowledgments**    I am indebted to Dr Luca Turing and Professor Jim Horne for valuable comments and references to other work. They are in no way responsible for any eccentricities of view of mine. These ideas emerged somewhat unexpectedly at the 2007 Quantum Mind conference, and I am indebted to the organisers for inviting me to a meeting whose range was perhaps even wider than they had intended.